\documentclass{PoS}

\newcommand{\met}       {\mbox{\ensuremath{\slash\kern-.7emE_{T}}}}

\newcommand{\ttbar}     {\mbox{$t\bar{t}$}}

\newcommand{\ppbar}     {\mbox{$p\bar{p}$}}

\newcommand{\comphep}   {\sc{c}\rm{omp}\sc{hep}}
\newcommand{\singletop} {\sc{singletop}}

\newcommand{\pythia}    {\sc{pythia}}
\newcommand{\alpgen}    {\sc{alpgen}}

\title{Observation of Single Top Quark Production with the D0 Detector}

\ShortTitle{D0 Single Top Observation}

\author{\speaker{Reinhard Schwienhorst}\thanks{On behalf of the D0 collaboration.}\\
	Michigan State University\\ East Lansing, MI\\ USA\\
        E-mail: \email{schwier@pa.msu.edu}}

\abstract{We report on the observation of single top quark production by the D0 collaboration
using a dataset of 2.3~fb$^{-1}$ collected at the Fermilab Tevatron \ppbar~collider. Several 
multivariate techniques are combined to separate the single top signal from backgrounds. The
measured single top cross section is
$\sigma_{tb+tqb}=3.94 \pm 0.88$~pb. The
probability to measure a cross section at this value or higher in the
absence of signal is $2.9\times10^{-7}$, corresponding to a
$5.0$~standard deviation significance for the presence of signal. 
The lower limit at the $95\%$~C.L. on the CKM matrix
element $V_{tb}$ is $|V_{tb}| > 0.78$. A separate measurement of the $t$-channel cross
section gives $\sigma_{tqb}=3.14^{+0.94}_{-0.80}{\rm ~pb.}$
}

\FullConference{European Physical Society Europhysics Conference on High Energy Physics,
                EPS-HEP 2009,\\
	        July 16 - 22 2009\\
	        Krakow, Poland}

\begin{document}

%
%
Top quarks were first observed as pairs produced via the strong interaction at the 
Fermilab Tevatron Collider in
1995~\cite{top-obs-1995-cdf,top-obs-1995-d0}. Single top quark production proceeds
via the weak interaction and its production cross section provides a direct measurement of the
the quark mixing matrix element $|V_{tb}|$~\cite{singletop-vtb-jikia}. It also
serves as a probe of the $Wtb$ 
coupling~\cite{Chen:2005vr,dudko-boos,singletop-wtb-heinson,d0-singletop-wtb,Abazov:2009ky}
and is sensitive to several models of new physics~\cite{Tait:2000sh}.

In 2007, D0 presented the first evidence for single top quark
production and the first direct measurement of
$|V_{tb}|$ using 0.9~fb$^{-1}$ of Tevatron data at a center-of-mass energy of 
1.96~TeV~\cite{d0-prl-2007,d0-prd-2008}. Recently, the
CDF collaboration has also presented such evidence in 2.2~fb$^{-1}$ of
data~\cite{cdf-prl-2008}. Here we describe the observation
of a single top quark signal in 2.3~fb$^{-1}$ of data~\cite{singletop-obs-d0}.
The CDF collaboration has also reported observation of single top quark
production~\cite{Aaltonen:2009jj}.

Single top quark production proceeds via the $s$-channel production and decay of a virtual
$W$~boson ($tb$)
and the $t$-channel exchange of a virtual $W$~boson ($tqb$).
We search for both of these processes at once as well as the $t$-channel process alone.
The sum of their predicted cross sections is $3.46 \pm 0.18$~pb~\cite{singletop-xsec-kidonakis} 
for a top quark mass $m_t = 170$~GeV. 

%
%

%
%
We select events collected with the D0 detector~\cite{NIM} containing
one isolated lepton (electron or muon), missing transverse energy,
and two, three, or four jets, at least one of which is
$b$-tagged. We separate the analysis into 24 channels by lepton type,
jet and $b$-tag multiplicity, and run period.
~
~
We model the signal using the {\comphep}-based next-to-leading order
(NLO) Monte Carlo (MC) event generator {\singletop}~\cite{singletop-mcgen}.
The main backgrounds to the single top final state signature are $W$+jets and $Z$+jets
production, as well as {\ttbar}, all of which are modeled using {\alpgen}~\cite{alpgen}.
A smaller background is from multijet events which are modeled using data
samples. Other backgrounds are from diboson production, modeled using 
{\pythia}~\cite{pythia}.

Systematic uncertainties arise mainly from the jet energy scale corrections 
and $b$-tag modeling, with smaller contributions from
MC statistics, correction for jet-flavor composition in $W$+jets events, 
and from the $W$+jets, multijets, and
{\ttbar} normalizations. The total uncertainty on the background is
(8--16)\% depending on the analysis channel, and we take both normalization
and shape effects into account.

We select 4519 events with a background expectation of 4428 events 
and a single top signal expectation of 223 events. Since the signal
is small compared to the overwhelming background, we apply three
separate multivariate analysis techniques based on boosted decision trees
(BDT)~\cite{decision-trees,bdt-benitez,bdt-gillberg}, Bayesian neural
networks (BNN)~\cite{bayesianNNs,bnn-tanasijczuk}, and the matrix
element (ME) method~\cite{matrix-elements,me-pangilinan}
to extract the single top signal. We then combine these in a combination
BNN.
\begin{figure}[!h!tbp]
\begin{center}
\includegraphics[width=1.66in]{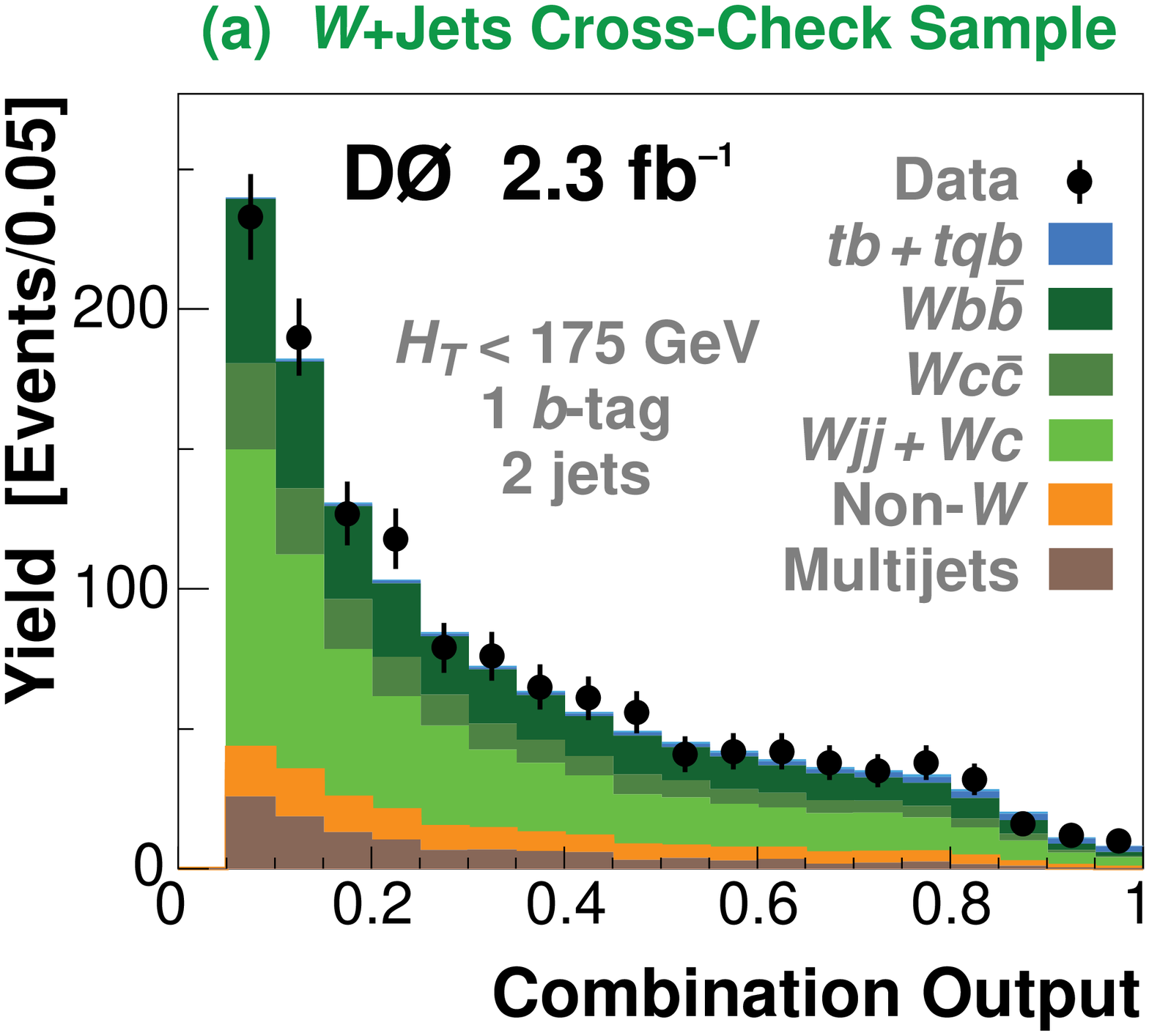}
\includegraphics[width=1.66in]{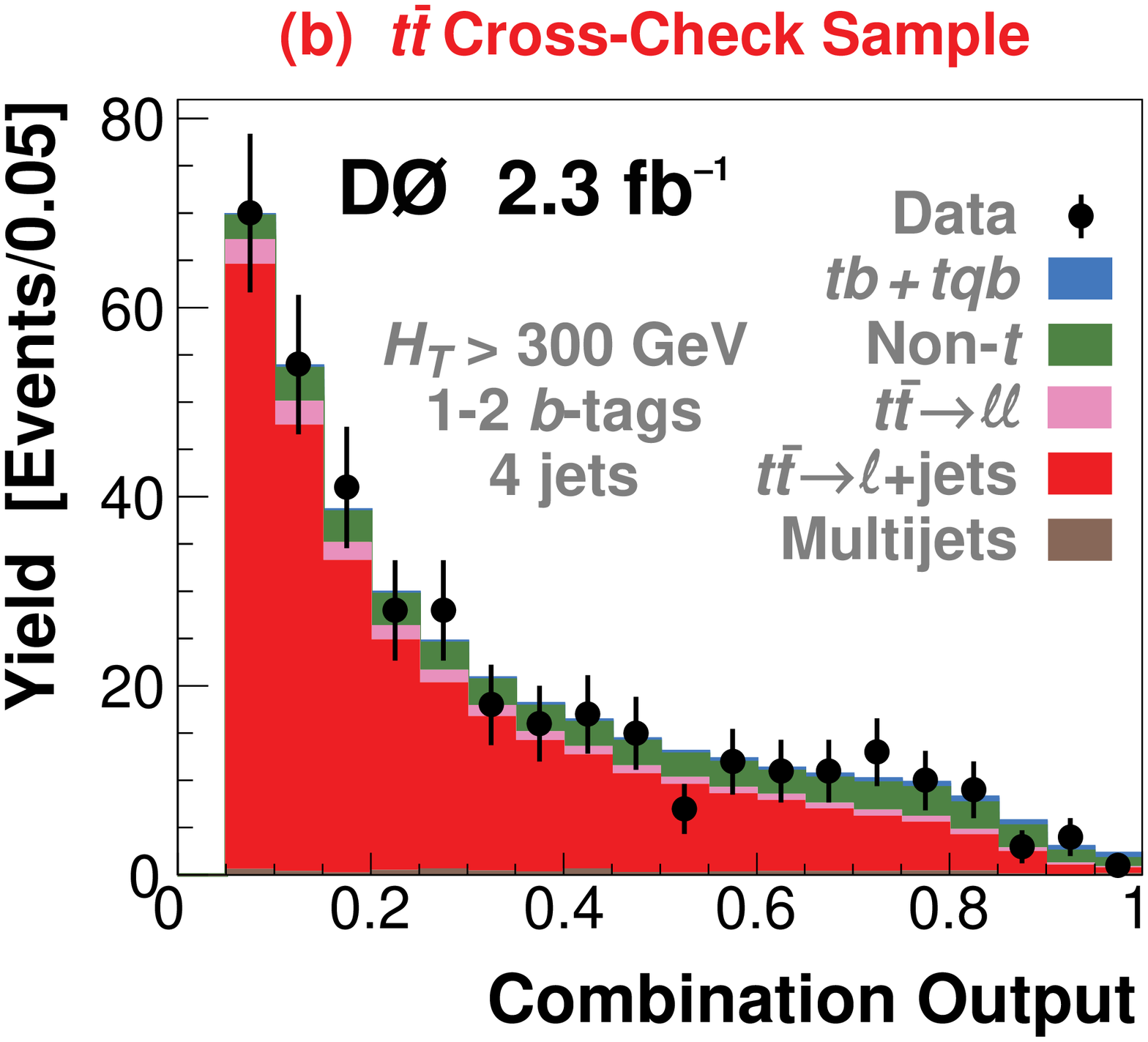}
\end{center}
\vspace{-0.22in}
\caption{Combination discriminant in two cross-check samples.}
\label{fig:cross-check}
\vspace{-0.15in}
\end{figure}

\begin{figure}[!h!tbp]
\begin{center}
\includegraphics[width=1.7in]{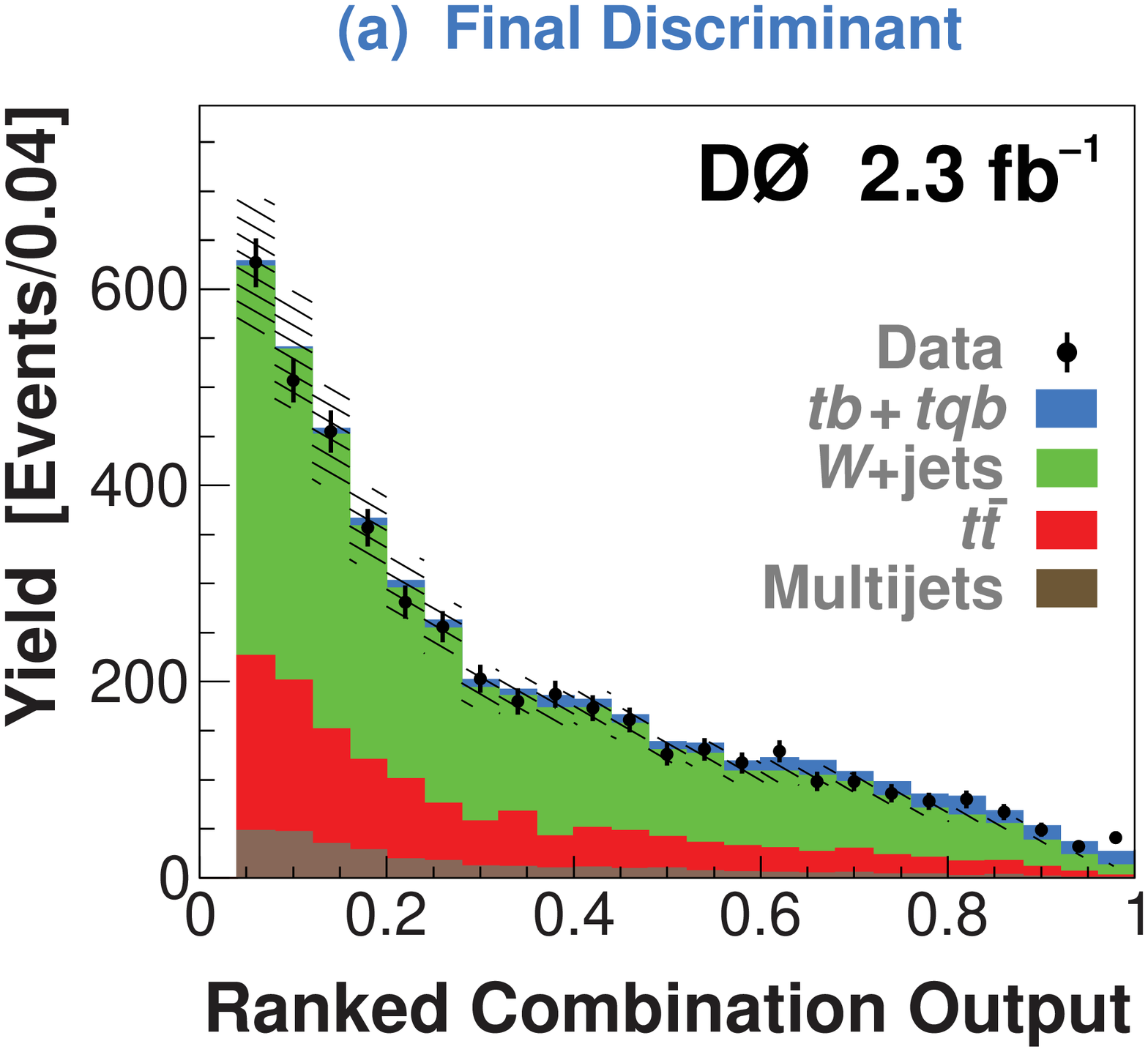}
\includegraphics[width=1.7in]{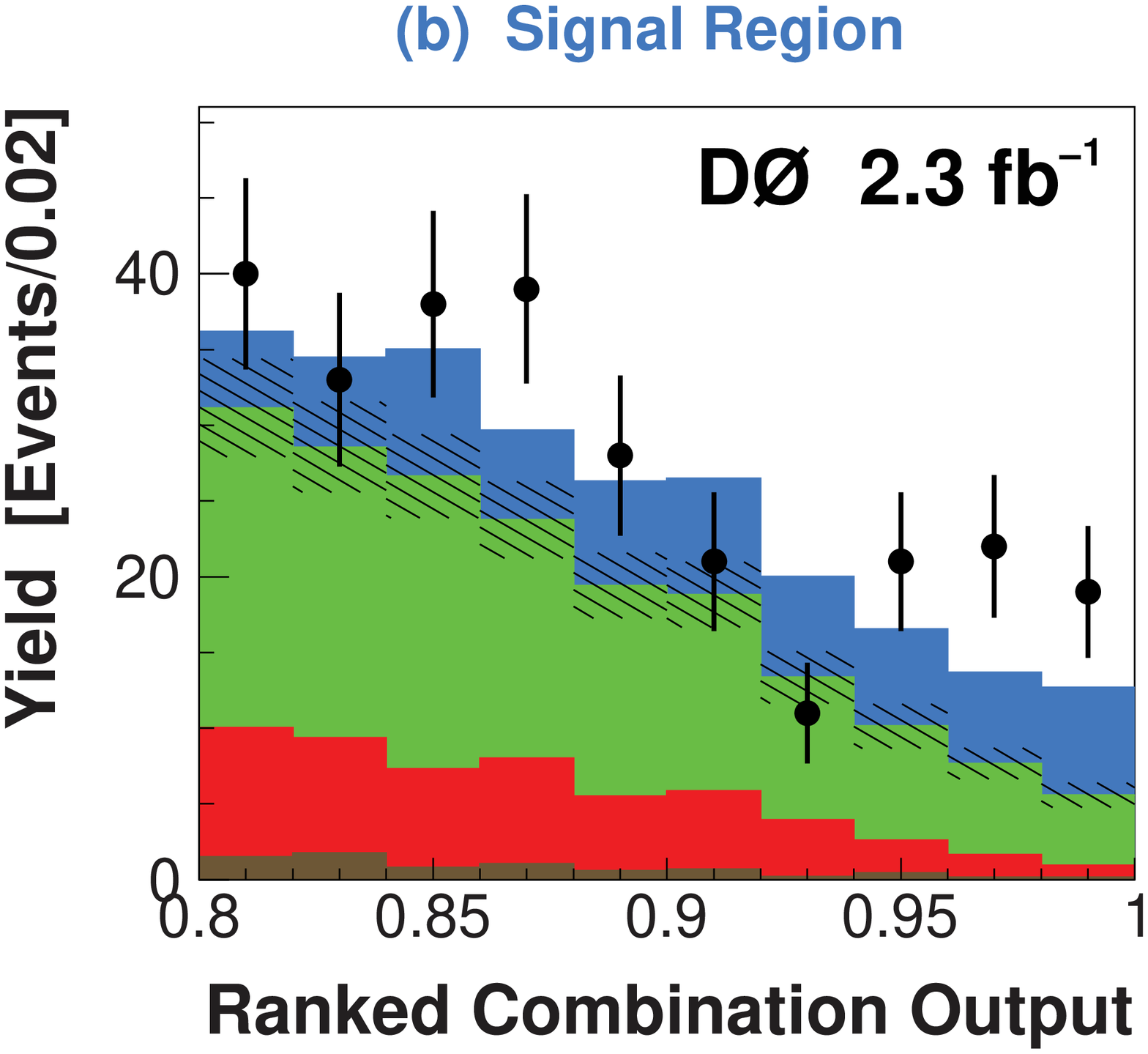}
\includegraphics[width=1.7in]{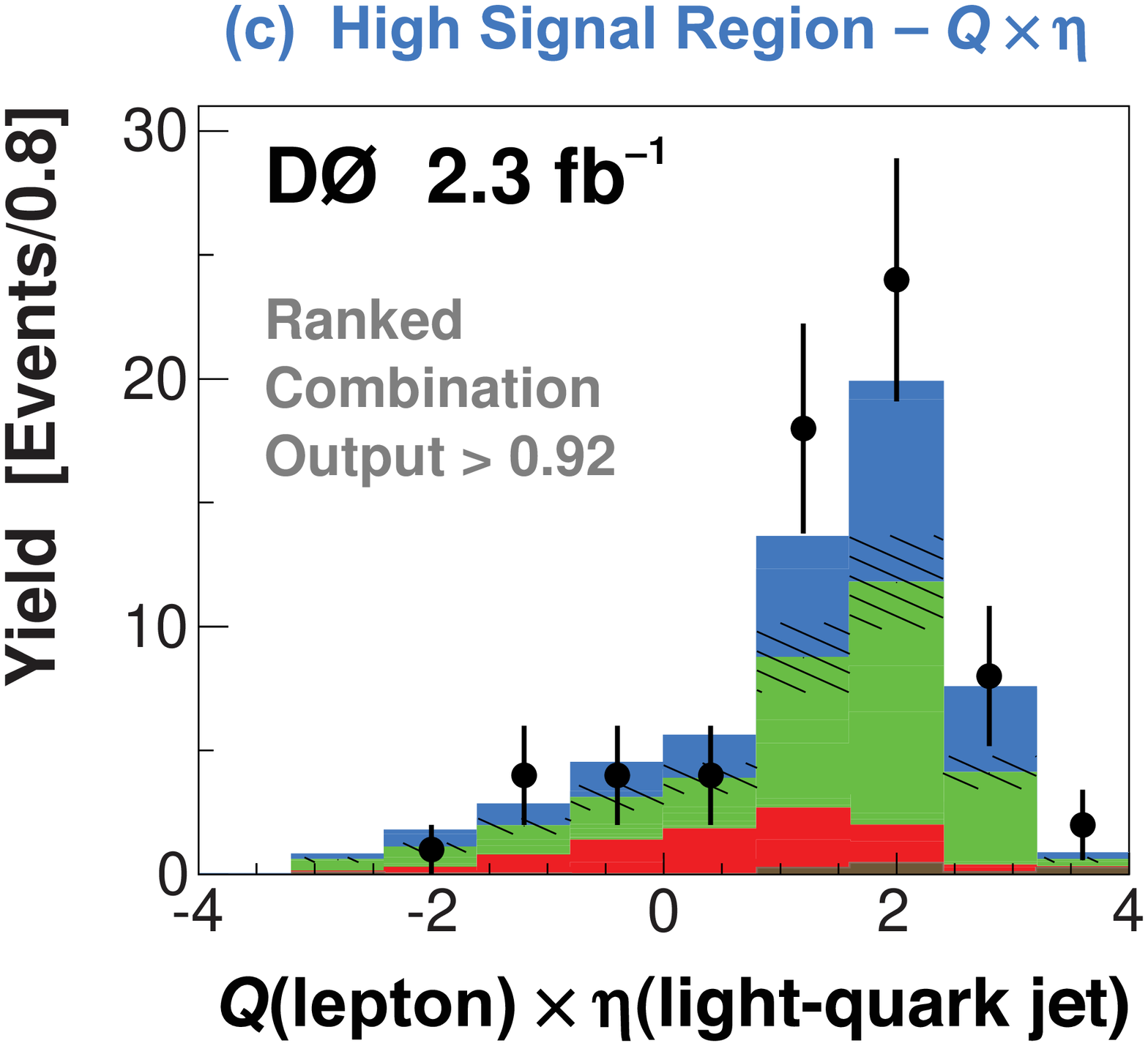}
\end{center}
\vspace{-0.22in}
\caption{Combination discriminant for all analysis
channels combined (a), only the signal region (b), and the pseudorapidity of the
light quark jet multiplied by the lepton charge for events in the signal region (c).
The signal is normalized to the measured cross section.}
\label{fig:combination}
\vspace{-0.15in}
\end{figure}

%
%
The BDT analysis uses a common set of 64 discriminating variables for all analysis channels,
but optimizes the filters separately in each channel. 
The BNN analysis selects 18--28 discriminating variables in
each channel. The ME analysis uses only 2-jet and 3-jet events and splits the analysis into
low-$H_T$ and high-$H_T$ regions at $H_T = 175$~GeV. All three analyses transform their
output distributions to ensure that every bin contains sufficient
background statistics. 
We verify the agreement between data and background model for each multivariate method
separate cross-check samples: a $W$+jets dominated sample and a \ttbar~dominated sample. The
combination discriminant output for these two samples is shown in Fig.~\ref{fig:cross-check}.
Fig.~\ref{fig:combination} shows the combination
discriminant output together with one of the discriminating variables for events in the
signal region. The measured cross section is
$\sigma_{tb+tqb} = 3.94 \pm 0.88 {\rm ~pb.}$
The measurement has a $p$-value of $2.6 \times 10^{-7}$,
corresponding to a significance of $5.0\,\sigma$. 
We use the cross section measurement to determine the Bayesian
posterior for $|V_{tb}|^2$ in the interval [0,1] and extract a limit
of $|V_{tb}| > 0.78$ at 95\% confidence level.

\begin{figure}[!h!tbp]
\includegraphics[width=1.8in]{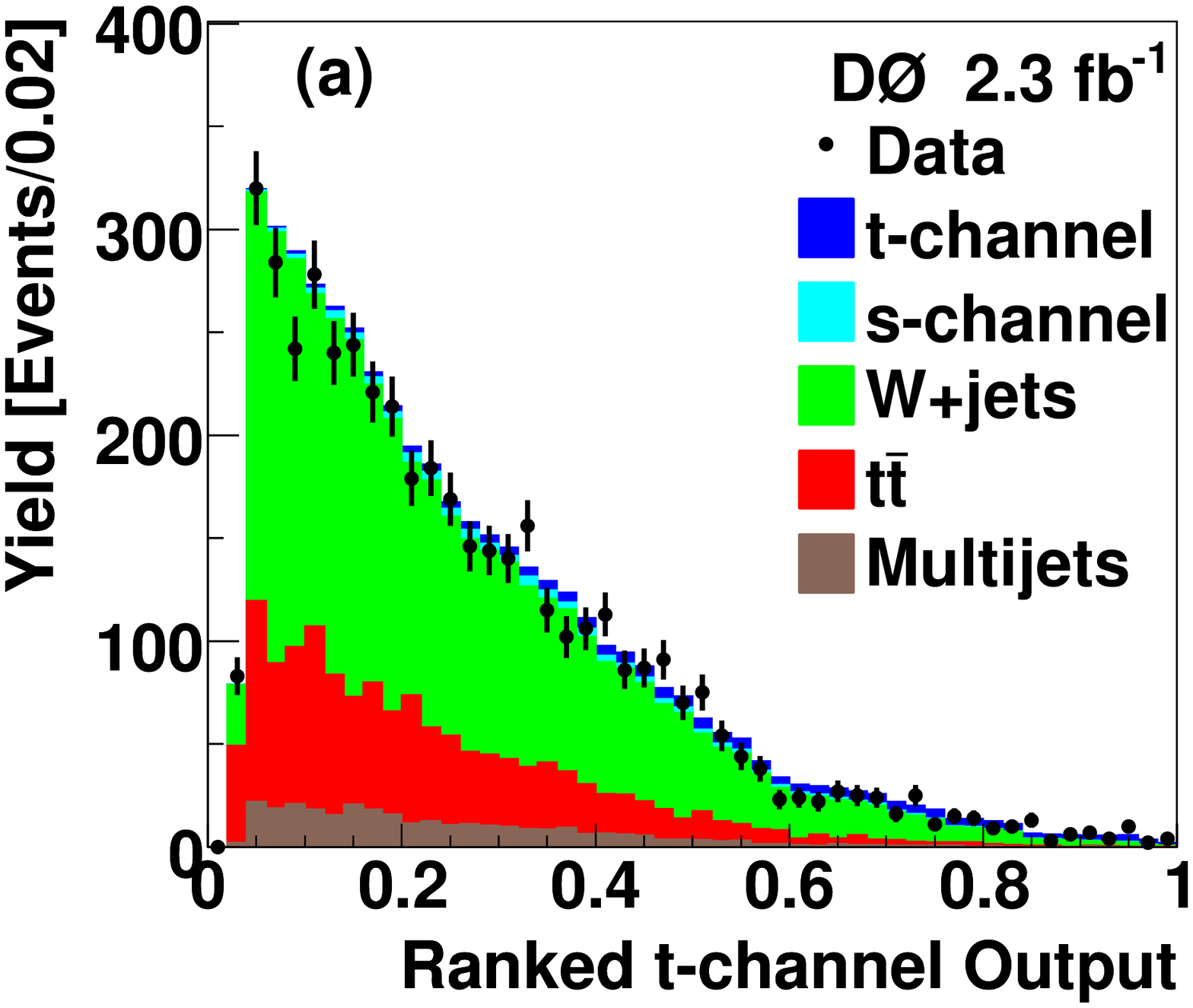} 
\vspace{-1.4in} \\
\includegraphics[width=1.8in]{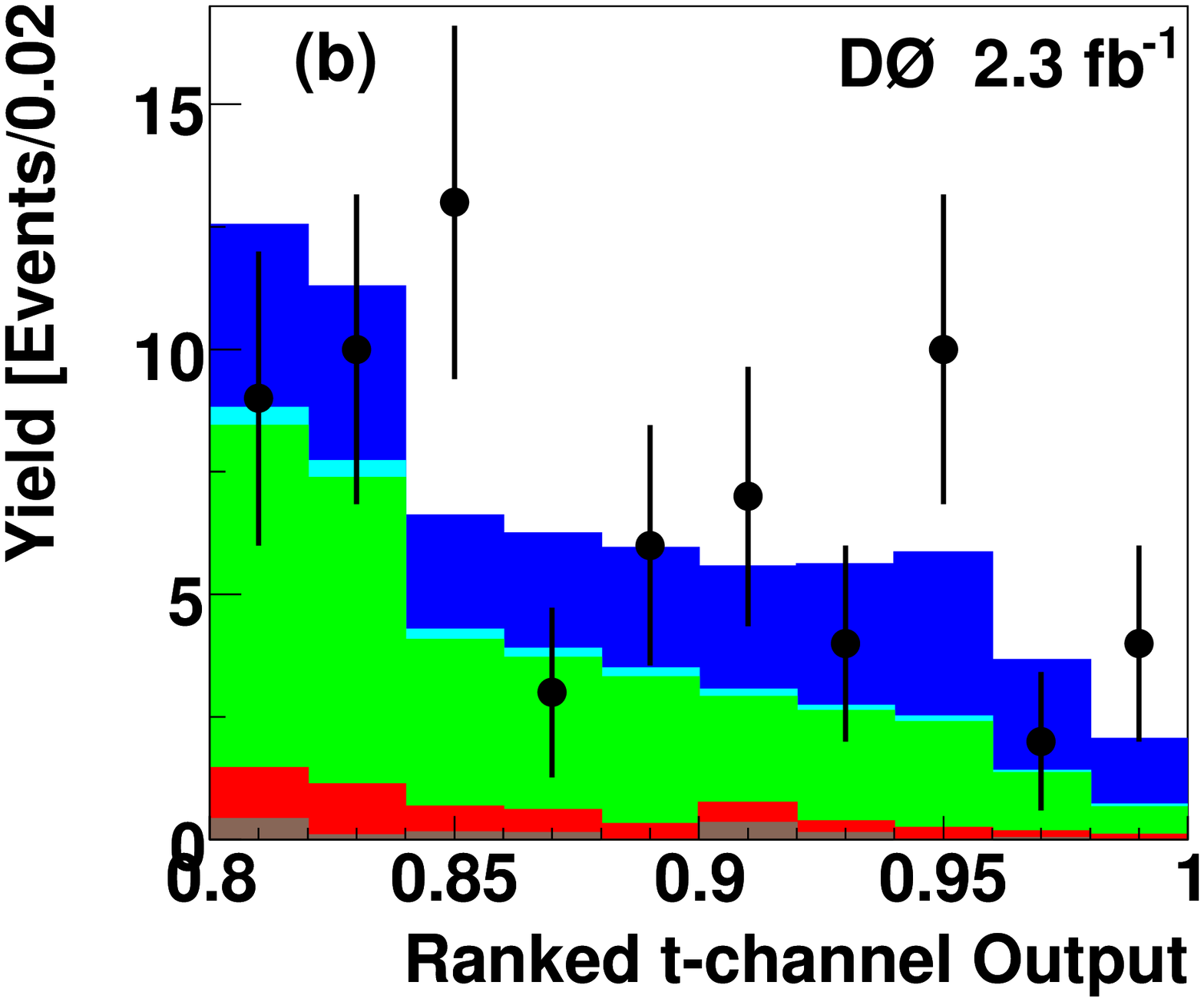}
\hspace{0.2in}
\includegraphics[width=3in]{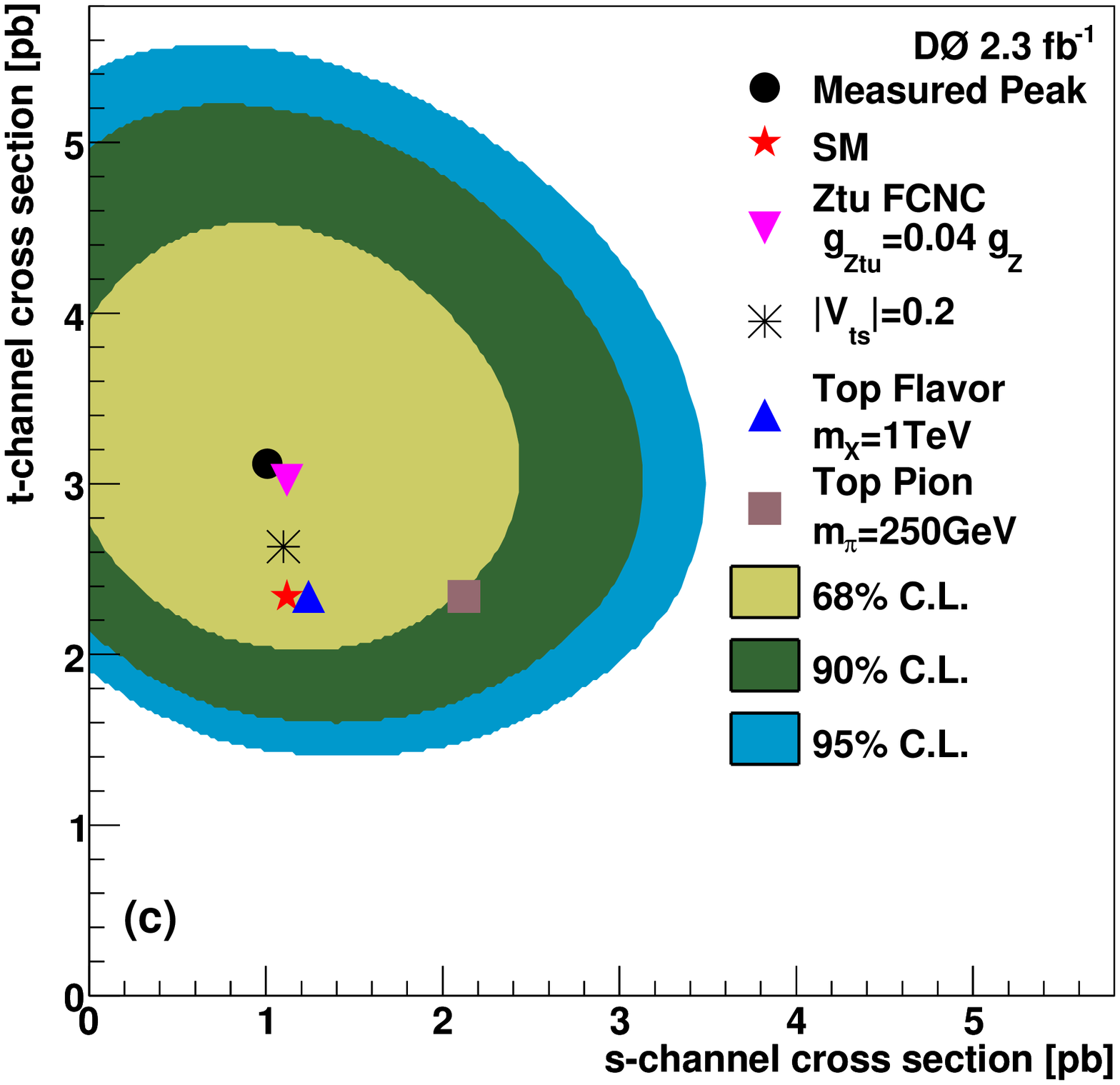}

\vspace{-0.12in}
\caption{Ranked $t$-channel discriminant (a), only the signal region (b), 
and  posterior probability density for $t$-channel and $s$-channel single top quark production
(c). Also shown are the measured cross section, SM
expectation, and several representative new physics scenarios~\cite{Tait:2000sh,Alwall:2007}.}
\label{fig:tchanpost}
\vspace{-0.15in}
\end{figure}
In a separate analysis we train the same multivariate methods using only $t$-channel
single top events as signal and including SM $s$-channel in the background~\cite{Abazov:2009pa}. 
The $t$-channel discriminant output is shown in Figs.~\ref{fig:tchanpost}(a) 
and~\ref{fig:tchanpost}(b).
We then determine the posterior as a function of both $s$-channel and $t$-channel
as shown in Fig.~\ref{fig:tchanpost}(c).
From this posterior we obtain the $t$-channel cross section by integrating over the
$s$-channel axis and find
$\sigma_{tqb} = 3.14^{+0.94}_{-0.80}{\rm ~pb.}$
We similarly extract the $s$-channel
cross section as $\sigma_{tb}=1.05 \pm 0.81$~pb by integrating over the $t$-channel axis.
The observed p-value for the $t$-channel measurement is $8.0\times 10^{-7}$,
corresponding to a Gaussian significance of $4.8\,\sigma$.

\end{document}